\documentclass[twocolumn]{aastex63}
\usepackage{amsmath}
\usepackage{graphicx}
\usepackage{rotating,tabularx}
\hypersetup{linkcolor=red,citecolor=blue,filecolor=cyan,urlcolor=black}

\defcitealias{Merlin2022}{Paper II}
\defcitealias{Castellano2022}{Paper III}
\defcitealias{Santini2022}{Paper XI}
\defcitealias{Treu2022b}{Paper XII}
\defcitealias{Yang2022}{Paper V}
\defcitealias{Leethochawalit2022}{Paper X}

\graphicspath{{./}{figures/}}

\shorttitle{The morphology of galaxies at the epoch of reionization}
\shortauthors{Treu et al.}

\begin{document}

\title{Early Results From  GLASS-JWST. XII: The Morphology of Galaxies at the Epoch of Reionization}


\correspondingauthor{Tommaso Treu}
\email{tt@astro.ucla.edu}

\author[0000-0002-8460-0390]{T.Treu}
\affiliation{Department of Physics and Astronomy, University of California, Los Angeles, 430 Portola Plaza, Los Angeles, CA 90095, USA}

\author[0000-0003-2536-1614]{A.~Calabr\`o}
\affiliation{INAF Osservatorio Astronomico di Roma, Via Frascati 33, 00078 Monteporzio Catone, Rome, Italy}

\author[0000-0001-9875-8263]{M.~Castellano}
\affiliation{INAF Osservatorio Astronomico di Roma, Via Frascati 33, 00078 Monteporzio Catone, Rome, Italy}

\author[0000-0003-4570-3159]{N.~Leethochawalit}
\affiliation{School of Physics, University of Melbourne, Parkville 3010, VIC, Australia}
\affiliation{ARC Centre of Excellence for All Sky Astrophysics in 3 Dimensions (ASTRO 3D), Australia}
\affiliation{National Astronomical Research Institute of Thailand (NARIT), Mae Rim, Chiang Mai, 50180, Thailand}

\author[0000-0001-6870-8900]{E.~Merlin}
\affiliation{INAF Osservatorio Astronomico di Roma, Via Frascati 33, 00078 Monteporzio Catone, Rome, Italy}

\author[0000-0003-3820-2823]{A.~Fontana}
\affiliation{INAF Osservatorio Astronomico di Roma, Via Frascati 33, 00078 Monteporzio Catone, Rome, Italy}

\author[0000-0002-8434-880X]{L.~Yang}
\affiliation{Kavli Institute for the Physics and Mathematics of the Universe, The University of Tokyo, Kashiwa, Japan 277-8583}

\author[0000-0002-8512-1404]{T.~Morishita}
\affil{Infrared Processing and Analysis Center, Caltech, 1200 E. California Blvd., Pasadena, CA 91125, USA}

\author[0000-0001-9391-305X]{M.~Trenti}
\affiliation{School of Physics, University of Melbourne, Parkville 3010, VIC, Australia}
\affiliation{ARC Centre of Excellence for All Sky Astrophysics in 3 Dimensions (ASTRO 3D), Australia}

\author[0000-0002-6317-0037]{A.~Dressler}
\affiliation{The Observatories, The Carnegie Institution for Science, 813 Santa Barbara St., Pasadena, CA 91101, USA}

\author[0000-0002-3407-1785]{C.~Mason}
\affiliation{Cosmic Dawn Center (DAWN)}
\affiliation{Niels Bohr Institute, University of Copenhagen, Jagtvej 128, DK-2200 Copenhagen N, Denmark}

\author{D.~Paris}
\affiliation{INAF Osservatorio Astronomico di Roma, Via Frascati 33, 00078 Monteporzio Catone, Rome, Italy}

\author[0000-0001-8940-6768]{L.~Pentericci}
\affiliation{INAF Osservatorio Astronomico di Roma, Via Frascati 33, 00078 Monteporzio Catone, Rome, Italy}

\author[0000-0002-4140-1367]{G.~Roberts-Borsani}
\affiliation{Department of Physics and Astronomy, University of California, Los Angeles, 430 Portola Plaza, Los Angeles, CA 90095, USA}

\author[0000-0003-0980-1499]{B.~Vulcani}
\affiliation{INAF Osservatorio Astronomico di Padova, vicolo dell'Osservatorio 5, 35122 Padova, Italy}

\author[0000-0003-4109-304X]{K.~Boyett}
\affiliation{School of Physics, University of Melbourne, Parkville 3010, VIC, Australia}
\affiliation{ARC Centre of Excellence for All Sky Astrophysics in 3 Dimensions (ASTRO 3D), Australia}

\author[0000-0001-5984-0395]{M.Bradac}
\affiliation{University of Ljubljana, Department of Mathematics and Physics, Jadranska ulica 19, SI-1000 Ljubljana, Slovenia}
\affiliation{Department of Physics and Astronomy, University of California Davis, 1 Shields Avenue, Davis, CA 95616, USA}

\author[0000-0002-3254-9044]{K.~Glazebrook}
\affiliation{Centre for Astrophysics and Supercomputing, Swinburne University of Technology, PO Box 218, Hawthorn, VIC 3122, Australia}

\author[0000-0001-5860-3419]{T.~Jones}
\affiliation{Department of Physics and Astronomy, University of California Davis, 1 Shields Avenue, Davis, CA 95616, USA}

\author[0000-0001-9002-3502]{D.~Marchesini}
\affiliation{Department of Physics and Astronomy, Tufts University, 574 Boston Ave., Medford, MA 02155, USA}

\author[0000-0002-9572-7813]{S.~Mascia}
\affiliation{INAF Osservatorio Astronomico di Roma, Via Frascati 33, 00078 Monteporzio Catone, Rome, Italy}

\author[0000-0003-2804-0648 ]{T.~Nanayakkara}
\affiliation{Centre for Astrophysics and Supercomputing, Swinburne University of Technology, PO Box 218, Hawthorn, VIC 3122, Australia}

\author[0000-0002-9334-8705]{P.~Santini}
\affiliation{INAF Osservatorio Astronomico di Roma, Via Frascati 33, 00078 Monteporzio Catone, Rome, Italy}


\author[0000-0002-6338-7295]{V.~Strait}
\affiliation{Cosmic Dawn Center (DAWN)}
\affiliation{Niels Bohr Institute, University of Copenhagen, Jagtvej 128, DK-2200 Copenhagen N, Denmark}

\author[0000-0002-5057-135X]{E.~Vanzella}
\affiliation{INAF -- OAS, Osservatorio di Astrofisica e Scienza dello Spazio di Bologna, via Gobetti 93/3, I-40129 Bologna, Italy}



\author[0000-0002-9373-3865]{X.~Wang}
\affil{Infrared Processing and Analysis Center, Caltech, 1200 E. California Blvd., Pasadena, CA 91125, USA}

\begin{abstract}

Star-forming galaxies can exhibit strong morphological differences between the rest-frame far-UV and optical, reflecting inhomogeneities in star-formation and dust attenuation. We exploit deep, high resolution NIRCAM 7-band observations to take a first look at the morphology of galaxies in the epoch of reionization ($z>7$), and its variation in the rest-frame wavelength range between Lyman $\alpha$ and 6000-4000\AA, at $z=7-12$. We find no dramatic variations in morphology with wavelength -- of the kind that would have overturned anything we have learned from the Hubble Space Telescope. No significant trends between morphology and wavelengths are detected using standard quantitative morphology statistics. We detect signatures of mergers/interactions in 4/19 galaxies. Our results are consistent with a scenario in which Lyman Break galaxies -- observed when the universe is only 400-800 Myrs old - are growing via a combination of rapid galaxy-scale star formation supplemented by accretion of star forming clumps and interactions.
\end{abstract}


\section{Introduction}
\label{sec:intro}

For more than a century it has been known that galaxies in the local universe do not come in every size, shape, and form \citep[e.g.,][]{hubble1926}. Rather, they can be easily classified into a small number of shapes, namely ellipticals, spirals, lenticulars, and irregulars. Galaxy morphology is the result of the underlying astrophysical process governing galaxy formation and evolution. A longstanding goal of extragalactic astronomy has been to explain why galaxies appear the way they do \citep[see][and references therein]{conselice14}. 

One of the major results of the Hubble Space Telescope (HST) has been that galaxy morphologies evolve with cosmic time, with classical elliptical and spiral galaxies dominating below $z\sim1$ and irregular and merging galaxies being more and more common at higher redshifts \citep[e.g.,][]{Lee13}. Structural properties and merger fractions have been investigated in the optical rest-frame up to redshift $z\sim3$ through imaging in the F814W, F105W, F125W, and F160W HST filters \citep{bond11,law12,vanderwel14,morishita14,huertascompany16,whitney21}. 

However, at redshift $z>3$ morphological studies become difficult. The optical rest frame shifts beyond the reach of HST's Wide Field Camera 3 (WFC3), forcing one to rely on rest frame UV light. Several works indeed extend the analysis of galaxy size and morphology at $z>3$ in this wavelength regime \citep[e.g.][]{conselice09,shibuya15,ribeiro16,bowler17}.
However, UV light is dominated by young stars and might therefore capture the location of star forming regions rather than the morphology of the bulk of stellar mass traced by more mature stars \citep[see, e.g.,][for a discussion of the effect]{rawat09}.
Furthermore, cosmological surface brightness dimming, limited angular resolution, and poor sampling of the WFC3-IR channel, degrade significantly even the available UV rest frame information. These effects, sometimes referred to as ``morphological'' K-correction \citep{kuchinski01,wuyts12}, can be substantial in terms of quantitative morphology.

Our morphological ignorance is particularly acute for galaxies at the epoch of reionization ($z\sim7$ and above). Only the far-UV rest frame is accessible to HST, and their sizes are so compact that they are typically only marginally resolved by WFC3 \citep[e.g.,][]{oesch10,Grazian12}. Strong lensing magnification helps with the angular resolution \citep[e.g.,][]{Yang22}, but cannot overcome the limitations in wavelength coverage. At last, with JWST we can overcome these limitations by virtue of its superior angular resolution and longer wavelength coverage with respect to HST.  

We use images obtained with NIRCam \citep{NIRCAM} on board the James Webb Space Telescope as part of the GLASS-JWST ERS program \citep{TreuGlass22} to take a first look at the morphology of $z>7$ galaxies. Our goal is to give a first answer to the questions ``What do galaxies at $z>7$ look like in the optical rest-frame?" and ``Are the UV and optical rest frame morphologies of z$>$7 galaxies similar or vastly different?", by applying well-established quantitative morphological methods to the revolutionary dataset. The 7-bands imaging dataset covers the observed wavelength from $0.8-4.8\mu$m, including the rest frame range between Lyman$\alpha$ and $\sim 4000$\AA\ up to $z\sim12$ (and up to $6000$\AA\ at $z\sim7$). The resolution (FWHM $0\farcs04$-$0\farcs14$) and sampling ($0\farcs031$-$0\farcs063$) are superior to HST in the overlapping regions and comparable all the way to the reddest band. Given the relatively small number of galaxies at $z>8.5$ in a single NIRCAM pointing, in this initial study we do not consider evolutionary effects above $z=7$, leaving the investigation of possible differences as a function of redshift to future work, based on larger samples. Two companion papers in the same focus issue discuss the size-luminosity relation of galaxies at $z>7$ \citep{Yang22b} and the morphology of galaxies after reionization is completed \citep{Jacobs2022}.

This letter is organized as follows. In Section~\ref{sec:sample} we summarize our sample selection. In Section~\ref{sec:methods} we summarize our quantitative morphological parameters. In Section~\ref{sec:morphology} we present our results. We discuss them in Section~\ref{sec:discussion}. We conclude in Section~\ref{sec:conclusions}.
Magnitudes are given in the AB system and a standard cosmology with $\Omega_{\rm m}=0.3, \Omega_{\rm \Lambda}=0.7$ and $h=0.7$ is assumed when necessary.

\section{Data and sample selection}
\label{sec:sample}

\begin{figure*}
\setcounter{figure}{0}
\center
\includegraphics[width=1\textwidth,trim={0.cm 29cm 0cm 0.cm},clip]{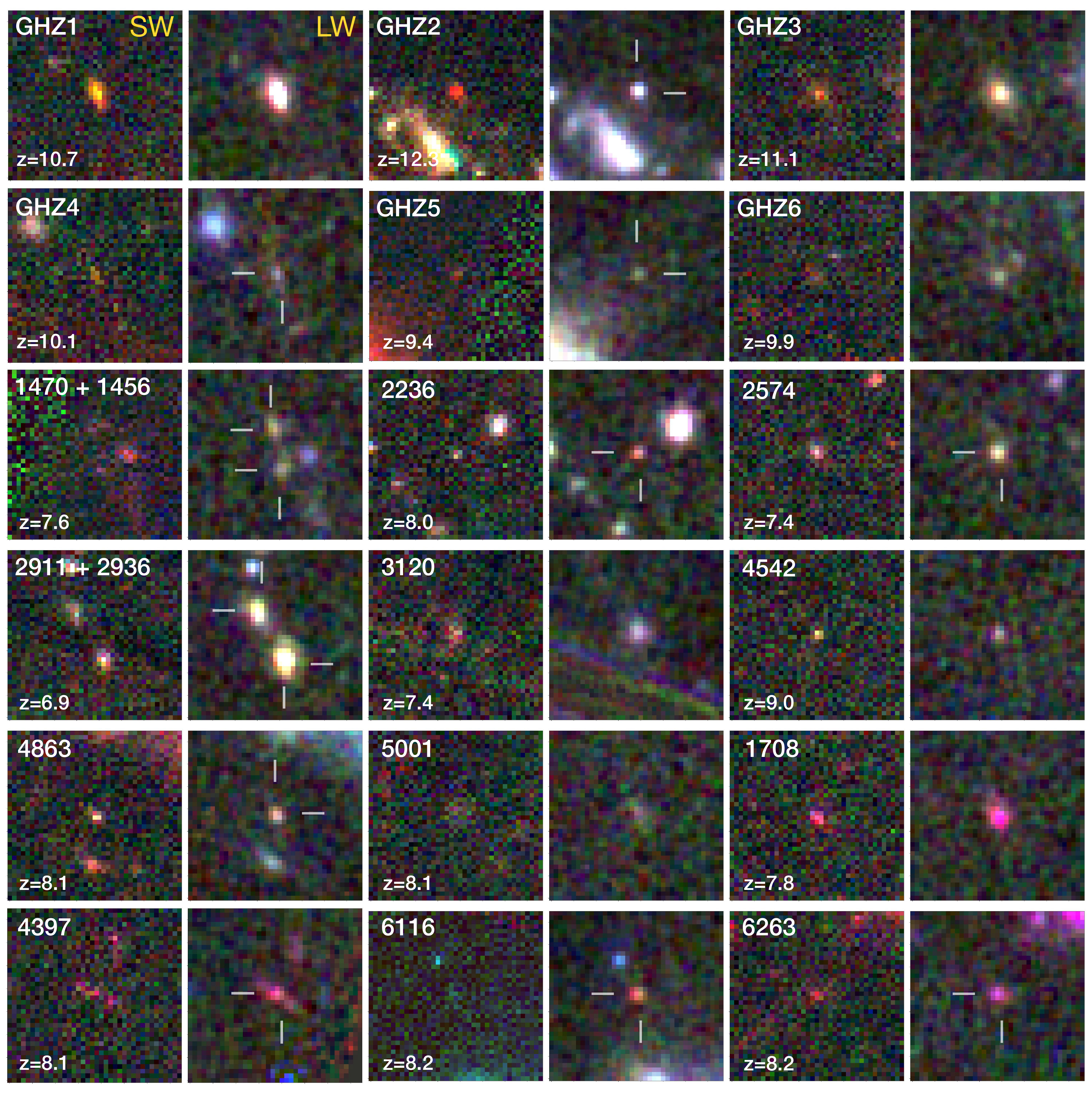}
\caption{For each galaxy we show a color composite image based on the short wavelength camera (B=F115W,G=F150W,R=F200W) and one based on the long wavelength camera (B=F277W,G=F356W,R=F444W). Individual images are degraded to the lower resolution of each camera (i.e., F200W and F444W, respectively). Postage stamps are $2.4\farcs$ on a side. Pixels are 31 mas and 63 mas respectively for the short and long wavelength images.}
\label{fig:galleryRGB}
\end{figure*}

We use NIRCam data obtained in parallel to NIRISS \citep{NIRISS} on June 28-29 2022. Details of the NIRCAM data quality, reduction, and photometric catalog creation can be found in the paper by \citet[][paper II]{Merlin2022}. Details of the NIRISS observations and data processing can be found in the paper by \citet[][paper I]{RobertsBorsani2022}.

After initial processing, samples of galaxies at $z>7$ are selected according to the ``drop-out" technique, as described by \citet[][paper X]{Leethochawalit2022} and \citet[][paper III]{Castellano2022}, supplemented by photometric redshifts. 
In total, our sample consists of $13$ galaxies selected from paper X and $6$ galaxies from paper III. The $19$ galaxies in our sample span a redshift range approximately from $7$ to $12$.

A color image gallery of our sample  is presented in Fig. \ref{fig:galleryRGB}, for each galaxy we show one color image based on the short wavelength camera using the F115W, F150W, and F200W bands, and one based on the three long wavelength channels, F270W, F356W, F444W. Image cutouts in each individual band are shown in Fig.~\ref{fig:gallery}. We do not show the F090W band because galaxies at this redshift are undetected owing to the opacity of the intergalactic medium. We notice that two galaxies of the sample presented by \citet{Leethochawalit2022} are likely interacting, so they are shown together in Fig. \ref{fig:galleryRGB} and Fig. \ref{fig:gallery}, and the morphological parameters are calculated  for the pair.s

\begin{figure*}[h!]
\setcounter{figure}{1}
\centering
\includegraphics[width=0.75\textwidth]{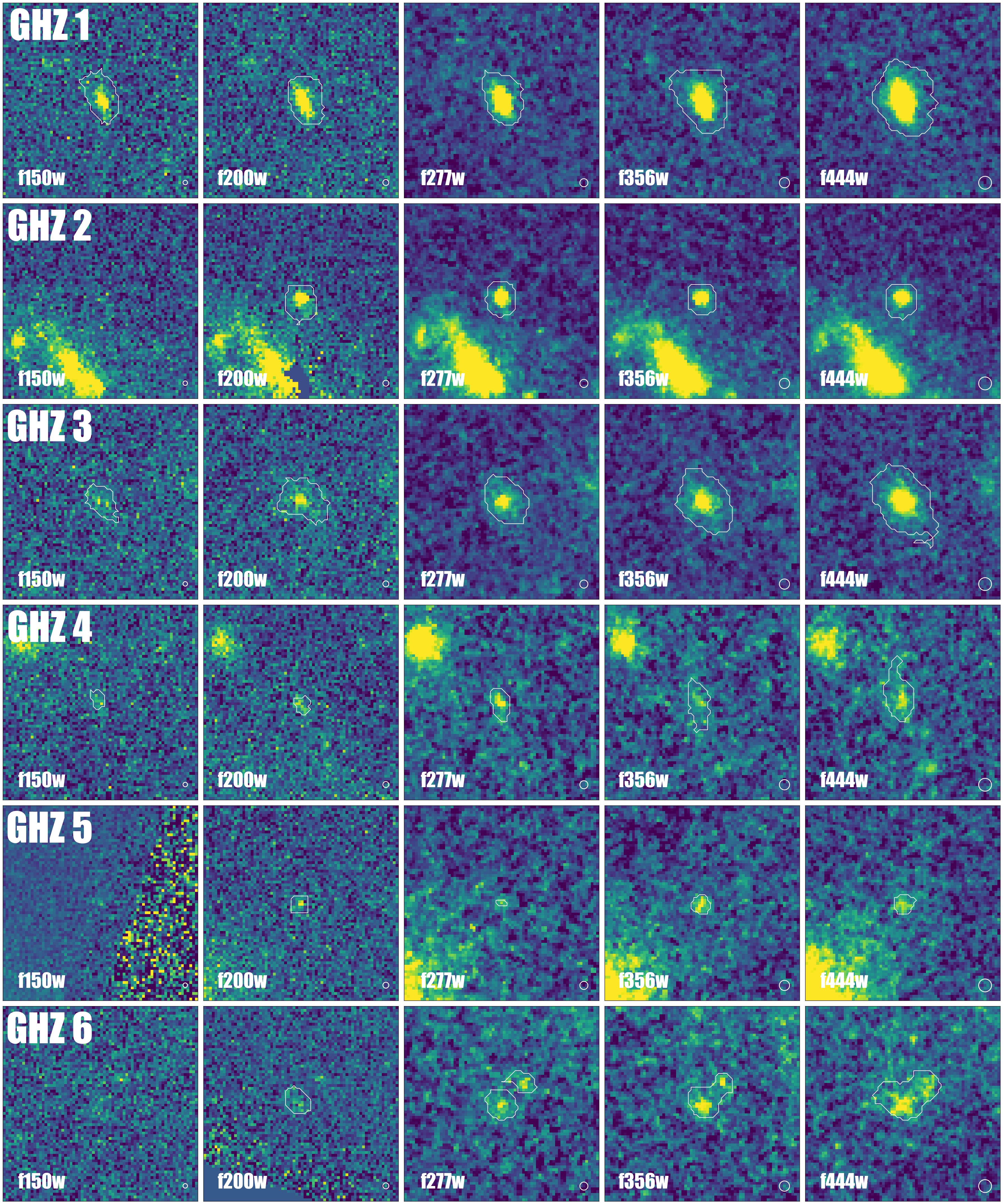}
\caption{Single band images of galaxies at $z>7$ selected by \citet[][Paper III]{Castellano2022} and \citet[][Paper X]{Leethochawalit2022}, in order of increasing wavelength of observation. The galaxies are identified by the ID used in papers III and X. Each postage stamp is $2.2\farcs$ on a side. The images are at their native resolution. Missing stamps are due to artefacts and edge effects. The white continuous lines delimitate the binary detection mask. The circles in the bottom right corner of each band are representative of the PSF FWHM size. In this first part of the figure are shown galaxies at $z\sim 9 - 12$ from paper III \citep{Castellano2022}.} 
\label{fig:gallery}
\end{figure*}

\begin{figure*}[h!]
\setcounter{figure}{1}
\centering
\includegraphics[width=\textwidth]{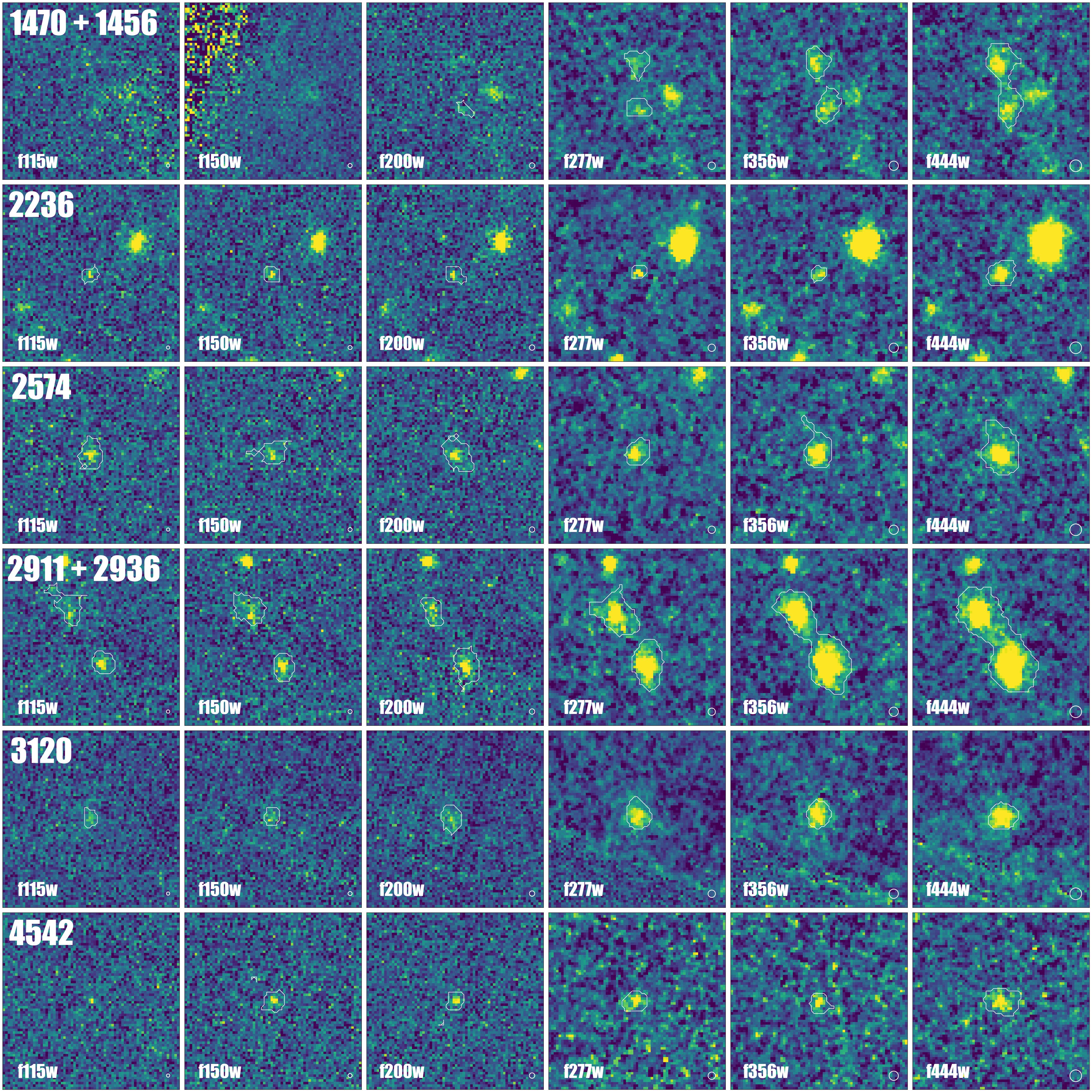}
\caption{Galaxies at $z\sim 7-9$ from paper X \citep{Leethochawalit2022}. Part 1.}
\end{figure*}

\begin{figure*}[h!]
\setcounter{figure}{1}
\centering
\includegraphics[width=\textwidth]{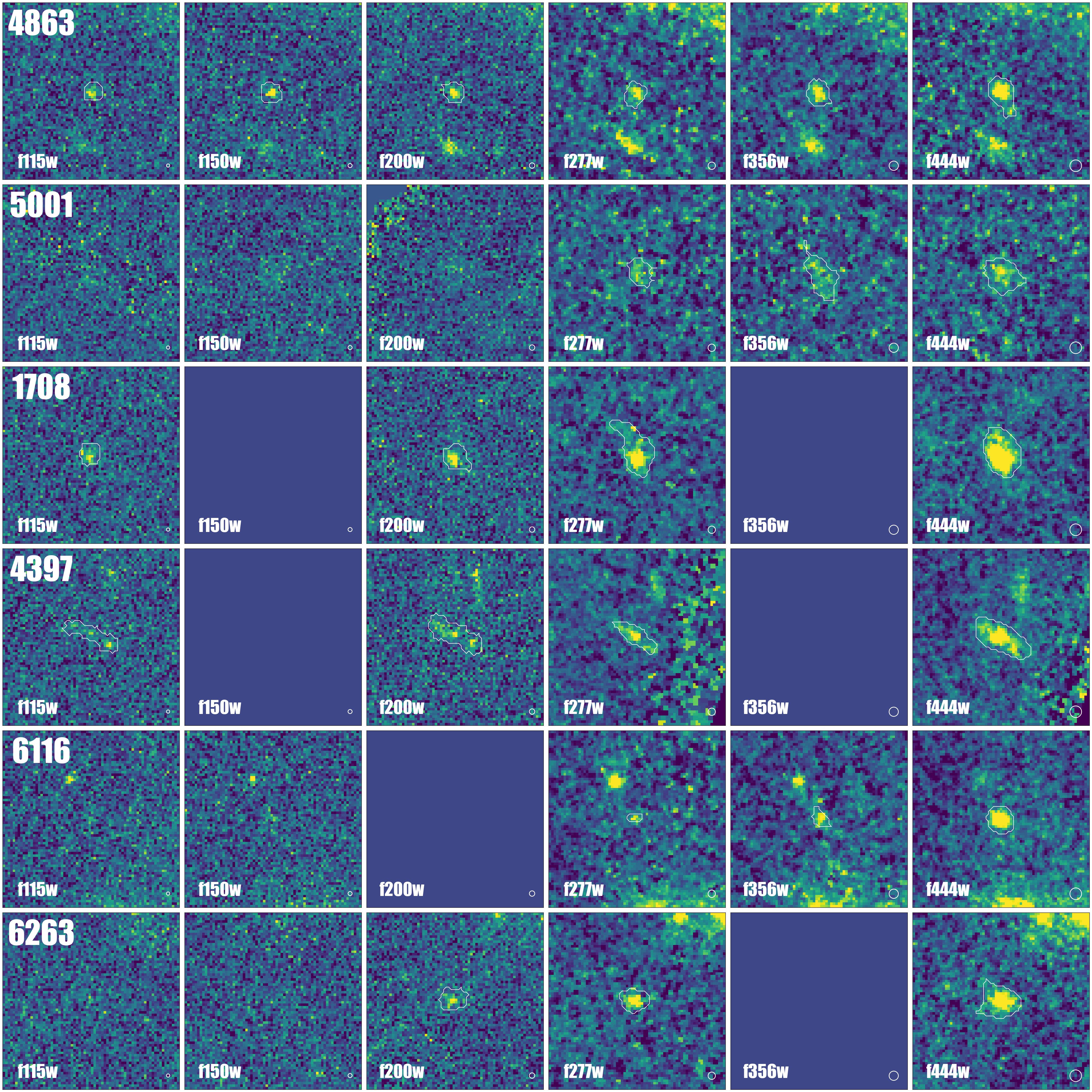}
Galaxies at $z\sim 7-9$ from paper X \citep{Leethochawalit2022}. Part 2.
\end{figure*}

\section{Methods}
\label{sec:methods}

\subsection{Definitions}

For each band, at its original resolution, we derive five well established quantitative morphological statistics according to the definitions introduced by previous works. The definitions are given below for convenience of the reader and to set the notation.\footnote{The codes used for the derivation of morphological parameters are fully accessible on the GitHub repository: \href{https://github.com/Anthony96/JWSTmorph.git}{https://github.com/Anthony96/JWSTmorph.git} }

First, we define a segmentation map for each object. Segmentation maps have been introduced in this context to reduce the impact of noise in the images and increase the signal from low surface-brightness regions. They are also recommended when dealing with the relatively low signal to noise expected for our galaxies \citep{pawlik16}. In practice, we apply a $6\times6$ uniform filter to the images and then derive a segmentation map with the \textit{photutils} astropy package, requiring for the sources at least $5$ connected pixels with a flux of $2 \sigma$ above the background. We finally define the binary detection mask of the object (M$_\text{D}$) as the segmentation region corresponding to our target (i.e. removing neighbors or non interacting companions). From M$_\text{D}$ we also derive the galaxy radius R$_\text{max}$ as the maximum pixel distance from the centroid of the binary detection mask (i.e., the pixel coordinates that minimize the shape asymmetry), which works better than the typical Petrosian radius in case of disturbed morphological shapes and low S/N \citep{pawlik16}. 

The Gini structural parameter (G) quantifies the degree of inequality of the light distribution in a galaxy \citep{abraham03,lotz04}, and is defined as:
\begin{equation}
G=\frac{1}{\bar Xn(n-1)}\sum_i^n(2i-n-1)X_i ,
\end{equation}\label{eq:gini}
where $n$ is the number of pixels assigned to the galaxy by the binary detection mask, X$_i$ are the intensities in each pixel $i$ (sorted in increasing order), and $\bar X$ is the mean pixel intensity. Gini ranges between $0$ (all the pixels have the same intensity) and $1$ (all the flux of the galaxy is concentrated in one pixel).

M$_{20}$ is defined as the normalized second order moment of the brightest $20\%$ pixels of the galaxy \citep{lotz04} :
\begin{equation}
\begin{aligned}
M_{20}=\log_{10}\left(\frac{\sum_i M_i}{M_{tot}}\right)\text{, with} \sum_i f_i < 0.2 f_{tot} \\
\text{with } M_{tot}=\sum_i^n M_i = \sum_i^n f_i [(x_i-x_c)^2+(y_i-y_c)^2]
\end{aligned}\label{eq:M20}
\end{equation}
where $x_i$ and $y_i$ are the pixel coordinates (inside the detection mask), while $x_c$ and $y_c$ correspond to the galaxy center where M$_\text{tot}$ is minimized. $f_i$ are the pixel intensities, while $f_\text{tot}$ is the total flux of the galaxy within R$_\text{max}$. 
This quantity increases with the number of off-centered bright features -- with typical values being in the range from $-3$ to $0$ -- and is usually anti-correlated with concentration. 

The concentration of light (C) is calculated as in \citet{pawlik16}:
\begin{equation}
C=5 \times \log_{10} \left( \frac{R_{80}}{R_{20}} \right),
\end{equation}\label{eq:concentration}
where R$_{80}$ and R$_{20}$ are the radii (from the same center used for R$_\text{max}$) enclosing $20\%$ and $80\%$ of the total galaxy flux defined above. 

\begin{figure*}
\setcounter{figure}{2}
\centering
 \includegraphics[angle=0,width=0.98\textwidth,trim={0.cm 1.cm 7.cm 0.cm},clip]{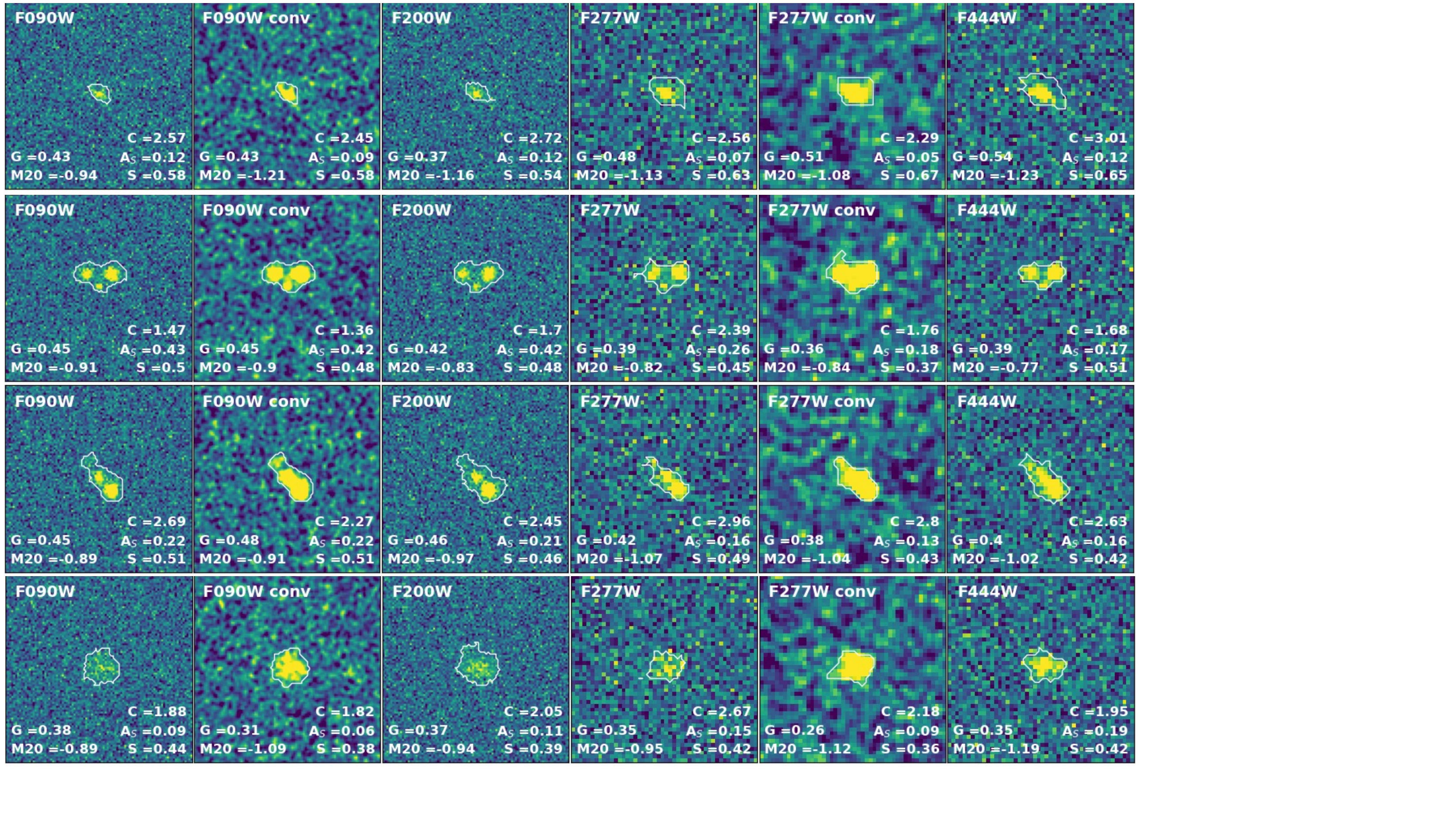}
 \caption{Examples of simulations carried out to estimate systematic errors arising from effects of sampling and correlated noise. A wide range of brightness and configuration were considered (see text), beyond those shown here. The stamps are $2"$ on a side. Images labeled ``F090W conv" and ``F277W conv" have been degraded to the resolution of the F200W and F444W band images, respectively.} 
 \label{fig:sims}
\end{figure*}

Shape asymmetry (A$_{\rm S}$) is defined by \citet{pawlik16} as :
\begin{equation}
A_S = \dfrac{\Sigma \mid M_\text{D}-M_{\pi} \mid}{2\ \Sigma M_\text{D}},
\end{equation}\label{eq:As}
i.e. the difference between the binary detection mask M$_\text{D}$ and the same mask rotated by $180$ degrees (dubbed M$_{\pi}$), summed over all the cutout pixels, and then divided by the number of pixels of the detection mask multiplied by $2$. The center of rotation is taken as the pixel coordinate for which A$_\text{S}$ is minimized. Compared to the standard definition of rotational asymmetry \citep{abraham96,conselice03}, A$_\text{S}$ is purely a measure of morphological asymmetry, regardless of the light distribution inside the galaxy. It is more sensitive to low surface brightness features \citep{pawlik16}, and therefore more appropriate for our goal of characterizing the shape of faint galaxies observed at $z\geq7$.

Finally, we also derive the smoothness parameter (S, sometimes called \textit{clumpiness}), which quantifies the contribution of small scale structures in a galaxy, as defined by \citet{conselice03}:
\begin{equation}
S = \dfrac{\Sigma \mid I_{0} - I_{\sigma} \mid}{\Sigma \mid I_{0} \mid} - S_{bkg}
\label{eq:S}
\end{equation}
where I$_0$ is the original galaxy image, $I_{\sigma}$ is derived by smoothing I$_0$ with a gaussian filter with $\sigma$ corresponding to a physical size of $1$ kpc in the source reference frame. 
Here, S$_\text{bkg}$ is the smoothness of the background. A completely smooth light distribution in a galaxy without bright small scale structures has S$=0$.
Owing to the compact nature of our sources, we do not attempt to identify a central nucleus or to remove the central galaxy regions in the computation of $S$. 

In order to minimize selection and evolutionary effects, we consider differential effects only, i.e. we focus on variations of each index with wavelength. In practice, for each galaxy and each index, we compute $\Delta$index as the difference between the measurement in a given filter and that obtained through F444W.

\subsection{Simulations}

In order to carry out a meaningful investigation of the variation of these statistics with wavelength, we have to estimate the amplitude of systematic uncertainties stemming from variations in resolution and pixel size across the NIRCAM bands, and from the effects of correlated noise when matching resolution. We proceed as follows. 

First, we simulate images in the F090W, F200W, F277W, and F444W bands, corresponding to the bluer and redder bands of the short wavelength (SW) and long wavelength (LW) channels, respectively.  We note that this is conservative, since in practice we do not use the F090W bands.

Second, we inject Gaussian sources in a set of configurations, aimed at sampling a range of shapes, geometry, size, and S/N, comparable to that of the expected real sources. In practice, we vary the total number of sources or clumps (from 2 to 4), their size (from $0.5$ to $2$ kpc at redshift 7), their relative position and brightness, the maximum angular extension of the configuration (from $0.2''$ to $0.5''$), and the total magnitude of the object (26, 27, and 28 AB). In addition, we simulate single sources with a Sersic profile, varying the Sersic index from $0.5$ to $4$, the ellipticity from $0$ to $1$, size and total magnitude as above. In total, we simulate $144$ $+$ $120$ different configurations. We also assume for simplicity that sources have a flat spectrum in $f_\nu$. For each band and configuration, we create $10$ images each with a different realization of the noise. Four of the configurations tested are displayed in each row of Fig.~\ref{fig:sims} as an illustration. 

Third, we create mock observations by adding shot noise and background noise, which is estimated in each band from the JWST ETC (v1.7), considering the integration times scheduled for our program \citep[i.e., $\sim12300$, $\sim5200$, $\sim5200$, and $\sim21000$ seconds from the bluer to the redder band;][]{TreuGlass22}. These correspond to the first, third, fourth, and sixth panel of each row in Fig. \ref{fig:sims}. 
In addition, we downgrade the resolution of our images by convolving them with a Gaussian kernel in order to match the PSF size to the redder bands (i.e., from $0.035''$ to $0.065''$ for the SW detector, and from $0.09''$ to $0.14''$ for the LW detector). 
This exercise allows us to determine how much the morphological parameters change when introducing correlated noise and lowering the resolution.  The convolved images are shown in Fig. \ref{fig:sims} in the second and fourth panel of each row. 

\begin{table}[h!]
\centering 
\renewcommand{\arraystretch}{1.5} 
\vspace{+0.1cm}
\begin{center} { \normalsize
\begin{tabular}{ | m{1cm} | m{0.8cm} | m{1cm} |m{0.8cm} | m{1.1cm} | m{0.6cm} |}
  \hline
  mag$_{AB}$ & $\Delta G$ & $\Delta$ M$_{20}$ & $\Delta$ C & $\Delta$ A$_S$ & $\Delta$ S  \\
  \hline
  26 & 0.02 & 0.05 & 0.2 & 0.015 & 0.02 \\ 
  27 & 0.025 & 0.06 & 0.3 & 0.020 & 0.02 \\ 
  28 & 0.03 & 0.08 & 0.4 & 0.030 & 0.02 \\ 
  \hline
\end{tabular} }
\end{center}

\caption{\small $1\sigma$ uncertainty on $\Delta$ index (index - index$_{F444W}$) for the five morphological parameters studied in this paper, as estimated from simulations.}
\label{tab:table_simulations}
\end{table}

Fourth, we measure morphological parameters for all the 
simulated images in each configuration, and compute differences with respect to the reddest band, as done for the observations. For each parameter we take the standard deviation as the estimated systematic uncertainty on $\Delta$index. 
We summarize in Table~\ref{tab:table_simulations} the uncertainties on all the parameters for three different values of the object total magnitude. We find that, as expected, systematic uncertainties increase with total magnitude on average, except smoothness, which remains constant. 

Consistent with previous findings by \citet{lotz04} for HST, we find that the indexes are systematically affected by the S/N of the images. In our simulations, we find that Gini increases with the average S/N per pixel estimated inside the segmentation maps of the galaxies. In contrast, M$_{20}$, S, A$_S$ and C decrease with S/N. In particular, the M$_{20}$, A$_S$, and C have a rapid variation below S/N $\sim2$, which makes the derivation of a correction factor rather difficult. Therefore, following \citet{lotz04}, we do not perform measurements when the average S/N per pixel is $<2$. In all cases, the parameters become stable above a S/N per pixel of $8$. We thus take the values at higher S/N per pixel as our `truth' value and then derive differential corrections (as a function of the average S/N per pixel) that we apply on the observed galaxies in each band. We note that we are interested in trends as a function of wavelength for each galaxy. Therefore we do not need to worry about comparing indexes for different galaxies with widely different total S/N.

\section{Results}
\label{sec:morphology}

\begin{figure*}[t!]
\setcounter{figure}{3}
 \includegraphics[width=0.47\textwidth]{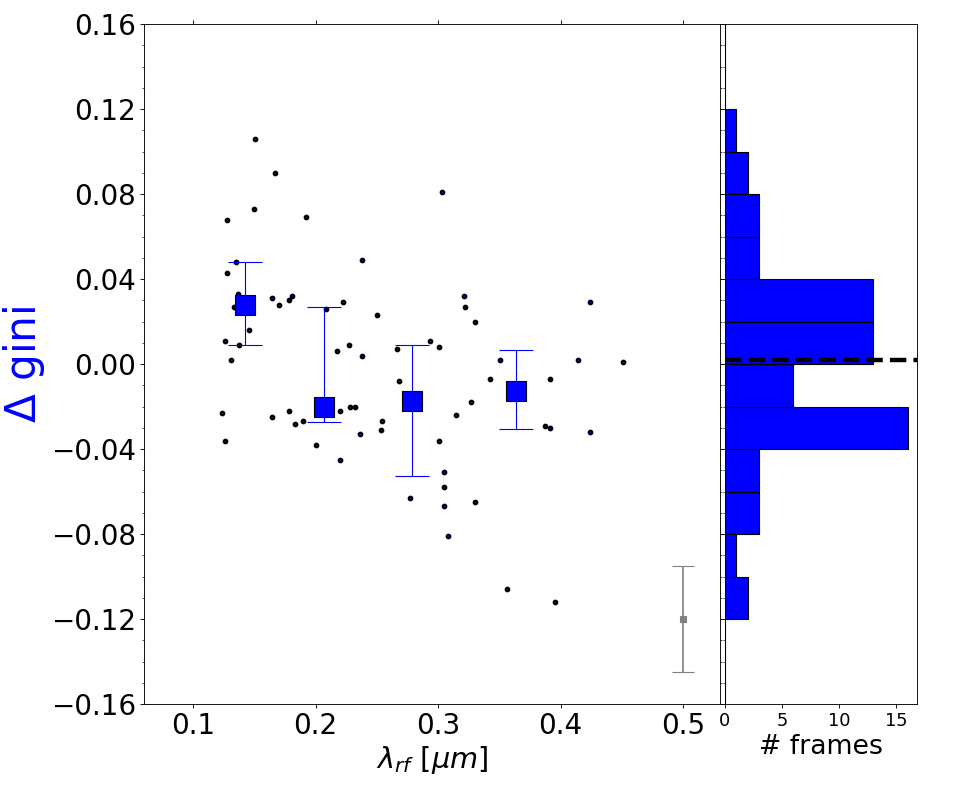}
 \includegraphics[width=0.47\textwidth]{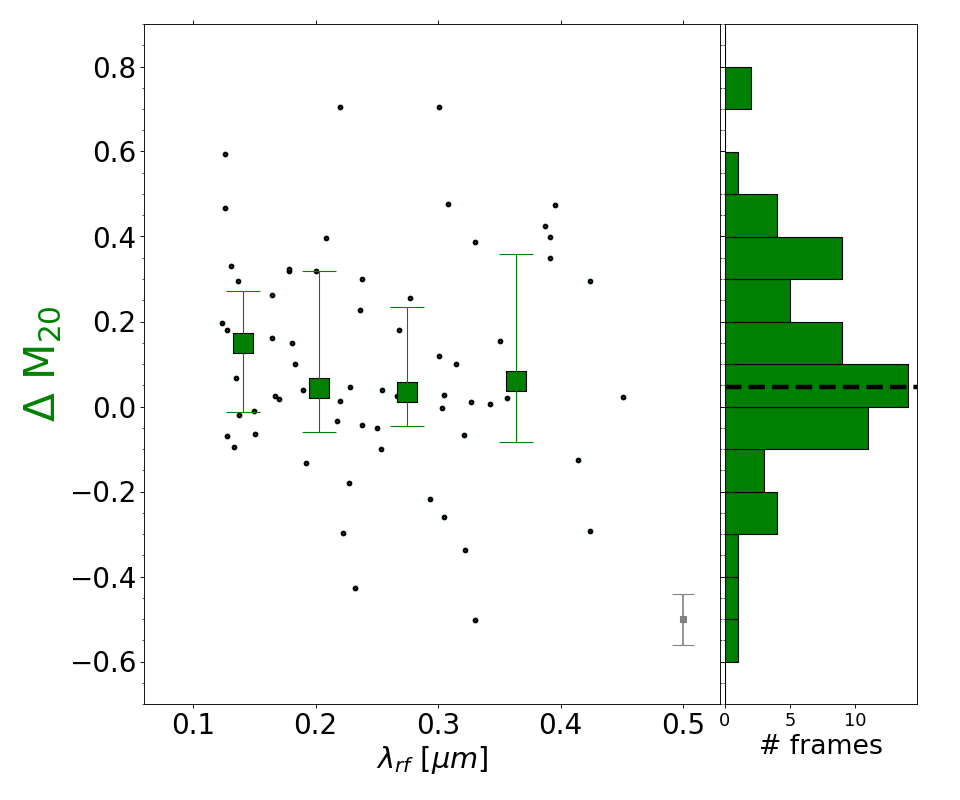}
 \includegraphics[width=0.47\textwidth]{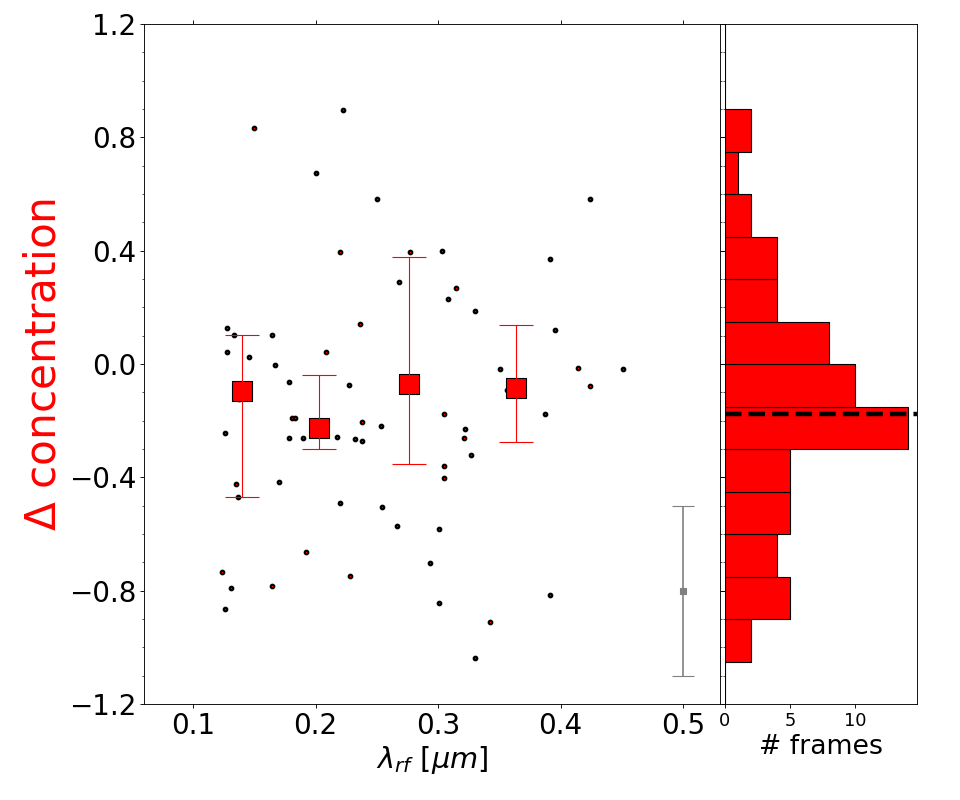}
 \includegraphics[width=0.47\textwidth]{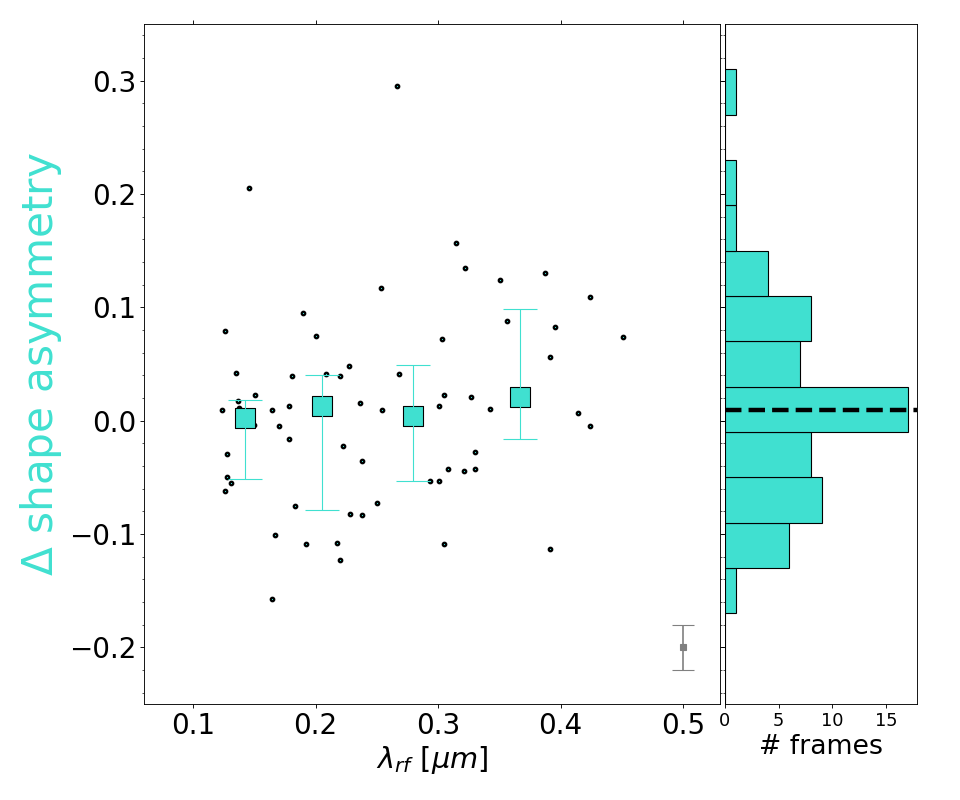}
 \includegraphics[width=0.47\textwidth]{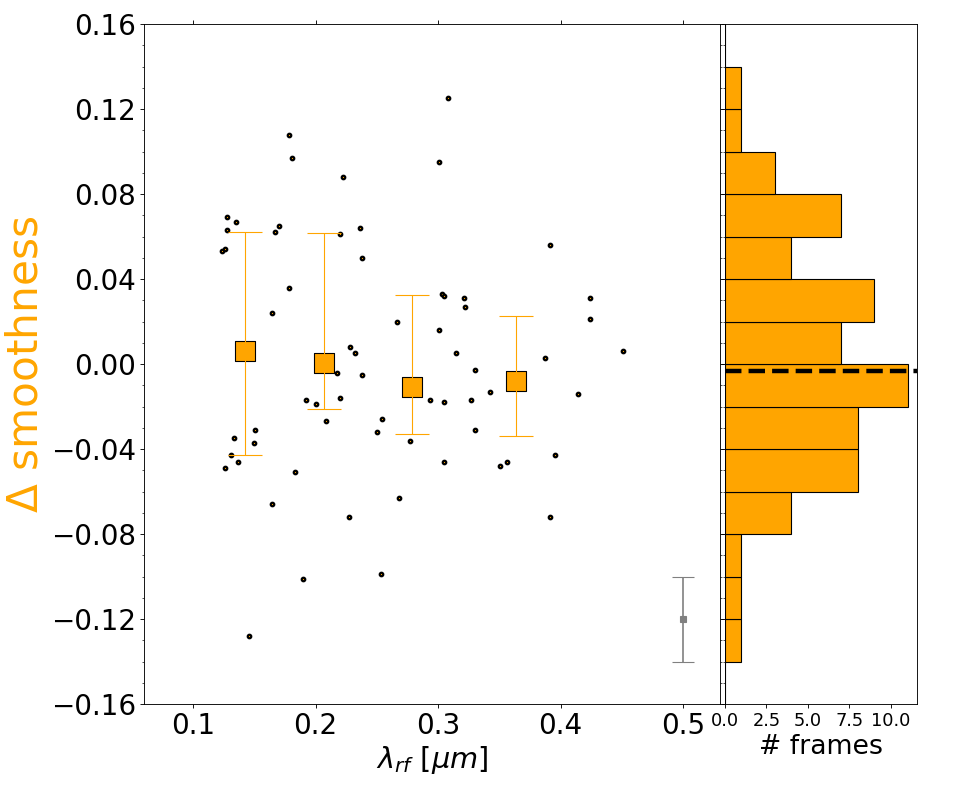} 
 \vspace{-0.3cm}
 \caption{Distribution of $\Delta$index ($=$ index $-$ index$_{F444W}$) for the five morphological parameters presented in the paper as a function of rest-frame wavelength. Each big square in the left plot of each panel represents the median value for the sample, while the error bars represent the first and third inter-quartiles of the distributions in each bin. The uncertainties on $\Delta$index are dominated by the systematic uncertainties summarized in Table~\ref{tab:table_simulations}, which are shown in the bottom right corner of each panel, at the median magnitude of the sample m$_{F444W} \simeq 27$. A histogram of $\Delta$index is shown in the right plot of each panel, with the median value as a dashed black line. }
 \label{fig:distributions}
\end{figure*}

We now investigate how the morphological parameters described in the previous section change across the six bands of the GLASS-JWST ERS survey (excluding F090W, where galaxies drop out), covering 1-5 $\mu m$. At our median redshift of $\sim8.4$, this allows us to probe the rest-frame range between Lyman $\alpha$ and $\sim5000$ \AA. For the most distant galaxy at $z\sim12$, the red filter includes light up to 3800\AA. 

Fig.~\ref{fig:distributions} summarizes our results. For each index we include a composite panel. The left part shows how the $\Delta$index (with respect to F444W) varies as a function of rest frame wavelength $\lambda_{\rm rf}$, while the right part shows the collapsed distribution.  In order to summarize the distributions in each band, 
we consider all the galaxies with reliable measurements (i.e. where S/N per pixel is above $2$), 
and compute both the median value of the $\Delta$index in four equally populated bins of $\lambda_{\rm rf}$, and then the first and the third interquartiles of its distribution.  
These quantities are drawn, respectively, with big colored squares and corresponding vertical error bars.

It is clear that, within our uncertainties and scatter, morphological indexes do not vary dramatically as a function of wavelength across the entire dynamic range probed by our observations. The median $\Delta$index is indeed in all cases consistent with zero within its $1\sigma$ uncertainty. 
However, in order to quantify the potential significance of any observed trends, we need to carry out a regression analysis, accounting for the uncertainties, estimated as in Section \ref{sec:methods}. 

Overall, we do not find significant variations of $\Delta$index as a function of wavelength, indicating that the properties of our galaxies across all the bands from far-UV to optical rest-frame do not change significantly. We also note that the distribution of $\Delta$index is approximately gaussian around the median values for all the parameters, and their FWHM are consistent with the level of uncertainty estimated from simulations for each $\Delta$index.

It is also worth looking at trends within individual galaxies, to identify systems where the indexes depend more strongly on wavelength, even though we expect that such trends would be more noisy compared to the previous average analysis.
For this purpose, we consider for each galaxy the slope of  $\Delta$index vs $\lambda_\text{rest-frame}$ with the available bands, and the median $\Delta$index. We find that the slope is not significantly different from $0$ on average, and the median varies by less than the systematic uncertainty. There are however a few exceptions where a significant dependency on wavelength is found. Galaxy ID $2911$, an interacting system, has a higher smoothness and lower shape asymmetry at longer wavelengths, while galaxy ID $2236$ has an opposite trend for the smoothness and shape asymmetry, and also a $>3 \sigma$ significant morphological diversity as a function of $\lambda$ of Gini and M$_{20}$. Finally, the scatter of $\Delta$index in individual galaxies is in general comparable to its uncertainty, and we do not see correlations among parameters in a same galaxy.

Furthermore, we also do not find in general a significant shift in the galaxy centroid between F444W and the bluest band available. 
However, we obtain small shifts (between $400$ and $800$ pc) for $\sim15 \%$ of the sample, which might be due in part to clumps having a different SED, as in the galaxy ID $4397$. 

Overall, we conclude that the variations of all the indexes from far-UV to optical rest-frame 
are certainly not dramatic, as we might have expected if the UV light corresponds to a small star forming region within a much larger galaxy as traced by older or dust obscured stars.

Finally, we analyze more in detail the effect of PSF smoothing. Deriving morphological parameters from images PSF-matched to the F444W band does not introduce significant differences for most of the indexes. The exceptions are Gini and concentration, for which we find that, while there is still no correlation with $\lambda_{rest}$ in Fig. \ref{fig:distributions}, $\Delta$index is systematically lower by $\sim0.04$ and $\sim0.4$, respectively, if they are measured on the smoothed images. The effect is also seen in our simulations.


\section{Discussion and comparison with previous work}
\label{sec:discussion}

\subsection{Morphology as a function of wavelength}\label{discussion1}

We have shown in the previous section that at $z>7$ morphological indexes do not vary significantly between UV and optical. This behavior is different from that reported by previous studies at $z\sim3-4$ and a far cry from the most extreme examples at lower redshift.

At $z\sim3-4$, \citet{Conselice08} and \citet{wuyts12} find that typical star-forming galaxies in the optical rest-frame have on average slightly smaller M$_{20}$, and a higher concentration and Gini coefficients than in the UV by $\sim 0.3$, $0.1$, and $0.3$, respectively, i.e., $1$ to $3$ times larger differences than our systematic uncertainties. This means that if similar variations are present at $z>7$, we would have been able to detect them. These trends at $z\sim3$-$4$ are explained as evidence for disk assembly through the inward migration of clumps and gas accretion. However, other similar studies report much milder or no morphological transformations with wavelength at $z\sim2.5$ \citep{dickinson99,papovich05}, which might be due in part to selection effects as their galaxies are bluer, with both UV and optical emission dominated by recent star formation. Similarly, \citet{bond11} claim that morphological differences between the rest-frame optical and UV in typical star-forming galaxies at $1.4< z < 3$ are small, which is likely due to uniform dust distributions.
As we discuss below, we believe that a version of these arguments - exacerbated by the extreme conditions at $z>7$ - is a possible explanation for our results.

In contrast, in the most extreme examples at low to intermediate redshift ($z \lesssim1$), larger differences with wavelength arise from inhomogeneity in the distribution of recent star formation \citep{rawat09,elmegreen09}, which typically occurs inside a disk with a larger scale length, and with the possible contribution from bright, off-centered clumps, or from patchy dust obscuration.

We now describe multiple factors that we believe contribute to the absence of strong wavelength dependency of the morphology. 

First, galaxies at $z=7-12$ have had very little time to form stars, since the universe is only 400-800 Myrs old at this point. Therefore, by necessity there cannot be much spectral  difference between the oldest stars and the ones that dominate the UV emission. Color trends are compressed by the timescale of the Universe. 

Second, the galaxies in our sample are vigorously star forming and not heavily dust obscured, owing to a combination of young ages and the Lyman Break selection technique \citep{Jaacks2018}.  Therefore, they did not have the time to build a substantial population of old stars, nor had the dust to hide a large fraction of young stars.

Third, in Lyman Break galaxies at these redshifts, star formation seems to be a global phenomenon, encompassing the majority of the galaxy, and not confined to a disk-like structure, or isolated star forming regions, like in the local universe.

Fourth, if star formation rate is smoothly rising as suggested by a number of authors \citep{finlator11}, most of the stars have recently formed, further reducing the time span available to give rise to morphological differences. For example, \citet{finlator11} predict that for this kind of SFH, optically and UV selected samples should be coincident, which would be consistent with the observed morphologically uniformity.

In conclusion, our results are qualitatively consistent with a scenario where LBGs at $z>7$ are growing via galaxy-scale star formation. Detailed analysis of larger samples, multiwavelength follow-up (especially with ALMA and with spectroscopy), is needed to reach a quantitative understanding of the relative contribution of these factors. This is left for future work.

Our results are consistent with those presented by \citet{Yang22b} in a companion paper. They carry out detailed surface photometry of the sample presented here, and they do not find any major variation in the size luminosity relation as a function of wavelength. This is at odds with the expectations of larger UV sizes coming from predictions of inside-out disk assembly models at high-redshift \citep{dutton11}. In turn, this suggests that the physical effects responsible for our patterns cannot be reconducted to the typical bulge - disk dichotomy emerging at lower redshift. 

A major caveat of our work, of course, is selection effects. In this first exploratory look, we have selected galaxies primarily following the Lyman Break technique. 
In some sense we have selected "normal" and common star forming galaxies.
This selection is heavily biased against dust enshrouded or quiescent galaxies. We do not know yet whether they exist at these redshifts. If they do, they will clearly not display the same amount of morphological regularity across wavelengths that we see in our sample.  Observations at even longer wavelengths, such as those with ALMA \citep{Inami2022}, will provide invaluable insights. Similarly, samples that are selected via emission lines, as opposed to the continuum, may reveal more morphological differences. JWST observations of samples selected in complementary ways may alter this first impression of morphological uniformity.

A second caveat is that these are just the first deep NIRCAM images, consisting of approximately 6 hours of exposure at F444W and less than 2 hours at F356W \citep{TreuGlass22}. Already the NIRSPEC parallels of GLASS-JWST will be deeper and it is not difficult to imagine integrating ten times as long in a deep field, once the instrument artifacts are properly understood \citep{Merlin2022}. Deeper images may reveal more morphological differences at lower surface brightness.

A third caveat is that these galaxies are extremely compact, with radii of just a few 100 pc (paper V). They did not have to be so compact in the optical, but since they are, there could be smaller scale morphological differences that are below even JWST resolution. Highly magnified sources will be valuable to overcome this limitation.

\subsection{Visual morphology and merger identification}\label{discussion2}

Consistent with the uniformity of morphological indexes as a function of wavelength, we also do not observe a dramatic change of the visual morphology of the galaxies in our sample (Figures~\ref{fig:galleryRGB} and~\ref{fig:gallery}). 

However, we notice that there is a diversity of shapes across the sample. Some of them are more compact, some of them are more elongated, and some of them have nearby companions, suggesting interactions or accretion of clumps. We leave a detailed exploration of the distribution of visual morphology to future work, when larger samples will be available at the end of the GLASS-JWST campaign. For the time being, we only comment on a few remarkable objects.

Four of our 19 galaxies are consistent with being interacting systems (21$\pm$10 \%; ID 6, 1470, 2911, and 2936). The first galaxy likely has a close and faint companion lying within its segmentation map. In the other two cases, the interactions are outside of the segmentation maps, thus we can actually check the photometric redshift of the companions. We also center the cutouts in the asymmetry center of the interacting systems in these cases. In the galaxy ID $1470$, we notice three closely separated objects with a relative distance of less than $1.0\farcs$ The two galaxies on the left have a photometric redshift of $7.6$ and $6.9$, consistent within their uncertainties, while the object on the right is likely an interloper with a z$_{phot}=0.55$. 
ID $2911$ is a bright interacting system at (photometric) redshift $\simeq 7$. 
We note that the fraction of interacting systems in our sample is similar to that of merging pairs identified by \citet{conselice09} among i-band dropouts at $z \sim6$ in the Hubble Ultra Deep Field.

One galaxy (ID $4397$) has a clumpy structure clearly visible in all the available bands, from F115W to F444W. 

The remaining galaxies of our sample appear instead isolated. In those few cases where a companion system is observed in the same cutout, 
we find that they are likely low-redshift interlopers. 

Statistical comparison with previous work at lower redshift requires a detailed assessment of incompleteness. It would be premature to carry out such study given our sample size. We leave this effort for future work, after the completion of the survey.  

\section{Conclusions}
\label{sec:conclusions}

JWST has given us new eyes to study the universe at $z>7$. For the first time, we have access to images of sufficient depth and resolution to characterize the morphology of galaxies deep into the reionization era, when galaxies were just 400-800 Myr old. In this letter, we have exploited NIRCAM data taken as part of the GLASS-JWST program to take a first look. The main results are as follows:

\begin{itemize}
    \item The morphology of Lyman Break Galaxies does not change significantly with wavelength, going from the rest frame optical to the rest frame UV. 
    \item  Four out of $19$ galaxies in our sample present clear signs of interaction or accretion.
\end{itemize}

We suggest a possible scenario that could at least qualitatively explain the observations. These galaxies are undergoing galaxy-scale rapid star formation. The timescales are extremely compressed given the young age of the Universe and the likely rising star formation rate \citep{finlator11}, leaving little time for the emergence of older stellar populations. Their compact sizes mean that crossing times are short (1-10 Myrs for speeds of order 10-100 km/s) compared to the spread in age of stellar populations that would be required to see a major difference, and therefore they are likely well mixed. Dust extinction is not sufficient in quantity or patchiness to induce detectable differences.  The clear detection of interacting systems is consistent with merging also contributing to the growth of these galaxies.

We conclude by listing some caveats that should be kept in mind and prevent us from drawing more quantitative conclusions at this time. First, our sample is small, and with a very clear selection function. Studies of galaxies selected at different wavelengths or through emission lines may reveal more morphological diversity as a function of wavelength. Second, the GLASS-JWST images are relatively deep but by no means the deepest that one can obtain with JWST. Already our second set of images will be significantly deeper. It is possible that deeper imaging may reveal lower surface brightness features that have escaped our detection. Third, these galaxies are compact, as shown in companion paper V, a few resolution elements across even with JWST. They did not have to be so compact in the optical rest frame -- and this is an important result -- but we cannot rule out more morphological diversity at even higher resolution. Studies of highly magnified galaxies will be very helpful in this respect. 

We plan to address some of the limitations identified here and carry out a detailed comparison with lower redshift work based on HST and JWST in the near future, after the completion of the GLASS-JWST observations.

\begin{acknowledgments}
  This work is based on observations made with the NASA/ESA/CSA James Webb Space Telescope. The data were obtained from the Mikulski Archive for Space Telescopes at the Space Telescope Science Institute, which is operated by the Association of Universities for Research in Astronomy, Inc., under NASA contract NAS 5-03127 for JWST. These observations are associated with program JWST-ERS-1324. We acknowledge financial support from NASA through grant JWST-ERS-1324.

  All the {\it JWST} data used in this paper can be found in MAST: \dataset[10.17909/fqaq-p393]{http://dx.doi.org/10.17909/fqaq-p393}.
  
KG acknowledges support from Australian Research Council Laureate Fellowship FL180100060. MB acknowledges support from the Slovenian national research agency ARRS through grant N1-0238. CM acknowledges support by the VILLUM FONDEN under grant 37459. The Cosmic Dawn Center (DAWN) is funded by the Danish National Research Foundation under grant DNRF140.
\end{acknowledgments}


\begin{thebibliography}{}
\expandafter\ifx\csname natexlab\endcsname\relax\def\natexlab#1{#1}\fi
\providecommand{\url}[1]{\href{#1}{#1}}
\providecommand{\dodoi}[1]{doi:~\href{http://doi.org/#1}{\nolinkurl{#1}}}
\providecommand{\doeprint}[1]{\href{http://ascl.net/#1}{\nolinkurl{http://ascl.net/#1}}}
\providecommand{\doarXiv}[1]{\href{https://arxiv.org/abs/#1}{\nolinkurl{https://arxiv.org/abs/#1}}}

\bibitem[{{Abraham} {et~al.}(1996){Abraham}, {Tanvir}, {Santiago}, {Ellis},
  {Glazebrook}, \& {van den Bergh}}]{abraham96}
{Abraham}, R.~G., {Tanvir}, N.~R., {Santiago}, B.~X., {et~al.} 1996, \mnras,
  279, L47, \dodoi{10.1093/mnras/279.3.L47}

\bibitem[{{Abraham} {et~al.}(2003){Abraham}, {van den Bergh}, \&
  {Nair}}]{abraham03}
{Abraham}, R.~G., {van den Bergh}, S., \& {Nair}, P. 2003, \apj, 588, 218,
  \dodoi{10.1086/373919}

\bibitem[{{Bond} {et~al.}(2011){Bond}, {Gawiser}, \& {Koekemoer}}]{bond11}
{Bond}, N.~A., {Gawiser}, E., \& {Koekemoer}, A.~M. 2011, \apj, 729, 48,
  \dodoi{10.1088/0004-637X/729/1/48}

\bibitem[{{Bowler} {et~al.}(2017){Bowler}, {Dunlop}, {McLure}, \&
  {McLeod}}]{bowler17}
{Bowler}, R.~A.~A., {Dunlop}, J.~S., {McLure}, R.~J., \& {McLeod}, D.~J. 2017,
  \mnras, 466, 3612, \dodoi{10.1093/mnras/stw3296}

\bibitem[{{Castellano} {et~al.}(2022){Castellano}, {Fontana}, {Treu},
  {Santini}, {Merlin}, {Leethochawalit}, {Trenti}, {Mestric}, {Vanzella},
  {Bonchi}, {Belfiori}, {Nonino}, {Paris}, {Polenta}, {Roberts-Borsani},
  {Boyett}, {Calabro}, {Glazebrook}, {Grillo}, {Mascia}, {Mason}, {Mercurio},
  {Morishita}, {Nanayakkara}, {Pentericci}, {Rosati}, {Vulcani}, {Wang}, \&
  {Yang}}]{Castellano2022}
{Castellano}, M., {Fontana}, A., {Treu}, T., {et~al.} 2022, ApJL, submitted,
  arXiv:2207.09436.
\newblock \doarXiv{2207.09436}

\bibitem[{{Conselice}(2014)}]{conselice14}
{Conselice}, C.~J. 2014, \araa, 52, 291,
  \dodoi{10.1146/annurev-astro-081913-040037}

\bibitem[{{Conselice} \& {Arnold}(2009)}]{conselice09}
{Conselice}, C.~J., \& {Arnold}, J. 2009, \mnras, 397, 208,
  \dodoi{10.1111/j.1365-2966.2009.14959.x}

\bibitem[{{Conselice} {et~al.}(2003){Conselice}, {Bershady}, {Dickinson}, \&
  {Papovich}}]{conselice03}
{Conselice}, C.~J., {Bershady}, M.~A., {Dickinson}, M., \& {Papovich}, C. 2003,
  \aj, 126, 1183, \dodoi{10.1086/377318}

\bibitem[{{Conselice} {et~al.}(2008){Conselice}, {Rajgor}, \&
  {Myers}}]{Conselice08}
{Conselice}, C.~J., {Rajgor}, S., \& {Myers}, R. 2008, \mnras, 386, 909,
  \dodoi{10.1111/j.1365-2966.2008.13069.x}

\bibitem[{{Dickinson}(1999)}]{dickinson99}
{Dickinson}, M. 1999, in American Institute of Physics Conference Series, Vol.
  470, After the Dark Ages: When Galaxies were Young (the Universe at 2 $<$ Z
  $<$ 5), ed. S.~{Holt} \& E.~{Smith}, 122--132, \dodoi{10.1063/1.58645}

\bibitem[{{Doyon} {et~al.}(2012){Doyon}, {Hutchings}, {Beaulieu}, {Albert},
  {Lafreni{\`e}re}, {Willott}, {Touahri}, {Rowlands}, {Maszkiewicz},
  {Fullerton}, {Volk}, {Martel}, {Chayer}, {Sivaramakrishnan}, {Abraham},
  {Ferrarese}, {Jayawardhana}, {Johnstone}, {Meyer}, {Pipher}, \&
  {Sawicki}}]{NIRISS}
{Doyon}, R., {Hutchings}, J.~B., {Beaulieu}, M., {et~al.} 2012, in Society of
  Photo-Optical Instrumentation Engineers (SPIE) Conference Series, Vol. 8442,
  Space Telescopes and Instrumentation 2012: Optical, Infrared, and Millimeter
  Wave, ed. M.~C. {Clampin}, G.~G. {Fazio}, H.~A. {MacEwen}, \& J.~{Oschmann},
  Jacobus~M., 84422R, \dodoi{10.1117/12.926578}

\bibitem[{{Dutton} {et~al.}(2011){Dutton}, {van den Bosch}, {Faber}, {Simard},
  {Kassin}, {Koo}, {Bundy}, {Huang}, {Weiner}, {Cooper}, {Newman}, {Mozena}, \&
  {Koekemoer}}]{dutton11}
{Dutton}, A.~A., {van den Bosch}, F.~C., {Faber}, S.~M., {et~al.} 2011, \mnras,
  410, 1660, \dodoi{10.1111/j.1365-2966.2010.17555.x}

\bibitem[{{Elmegreen} {et~al.}(2009){Elmegreen}, {Elmegreen}, {Fernandez}, \&
  {Lemonias}}]{elmegreen09}
{Elmegreen}, B.~G., {Elmegreen}, D.~M., {Fernandez}, M.~X., \& {Lemonias},
  J.~J. 2009, \apj, 692, 12, \dodoi{10.1088/0004-637X/692/1/12}

\bibitem[{{Finlator} {et~al.}(2011){Finlator}, {Oppenheimer}, \&
  {Dav{\'e}}}]{finlator11}
{Finlator}, K., {Oppenheimer}, B.~D., \& {Dav{\'e}}, R. 2011, \mnras, 410,
  1703, \dodoi{10.1111/j.1365-2966.2010.17554.x}

\bibitem[{{Grazian} {et~al.}(2012){Grazian}, {Castellano}, {Fontana},
  {Pentericci}, {Dunlop}, {McLure}, {Koekemoer}, {Dickinson}, {Faber},
  {Ferguson}, {Galametz}, {Giavalisco}, {Grogin}, {Hathi}, {Kocevski}, {Lai},
  {Newman}, \& {Vanzella}}]{Grazian12}
{Grazian}, A., {Castellano}, M., {Fontana}, A., {et~al.} 2012, \aap, 547, A51,
  \dodoi{10.1051/0004-6361/201219669}

\bibitem[{{Hubble}(1926)}]{hubble1926}
{Hubble}, E.~P. 1926, \apj, 64, 321, \dodoi{10.1086/143018}

\bibitem[{{Huertas-Company} {et~al.}(2016){Huertas-Company}, {Bernardi},
  {P{\'e}rez-Gonz{\'a}lez}, {Ashby}, {Barro}, {Conselice}, {Daddi}, {Dekel},
  {Dimauro}, {Faber}, {Grogin}, {Kartaltepe}, {Kocevski}, {Koekemoer}, {Koo},
  {Mei}, \& {Shankar}}]{huertascompany16}
{Huertas-Company}, M., {Bernardi}, M., {P{\'e}rez-Gonz{\'a}lez}, P.~G.,
  {et~al.} 2016, \mnras, 462, 4495, \dodoi{10.1093/mnras/stw1866}

\bibitem[{{Inami} {et~al.}(2022){Inami}, {Algera}, {Schouws}, {Sommovigo},
  {Bouwens}, {Smit}, {Stefanon}, {Bowler}, {Endsley}, {Ferrara}, {Oesch},
  {Stark}, {Aravena}, {Barrufet}, {da Cunha}, {Dayal}, {De Looze}, {Fudamoto},
  {Gonzalez}, {Graziani}, {Hodge}, {Hygate}, {Nanayakkara}, {Pallottini},
  {Riechers}, {Schneider}, {Topping}, \& {van der Werf}}]{Inami2022}
{Inami}, H., {Algera}, H., {Schouws}, S., {et~al.} 2022, \mnras,
  \dodoi{10.1093/mnras/stac1779}

\bibitem[{{Jaacks} {et~al.}(2018){Jaacks}, {Finkelstein}, \&
  {Bromm}}]{Jaacks2018}
{Jaacks}, J., {Finkelstein}, S.~L., \& {Bromm}, V. 2018, \mnras, 475, 3883,
  \dodoi{10.1093/mnras/sty049}

\bibitem[{{Jacobs} {et~al.}(2022){Jacobs}, {Glazebrook}, {Calabr{\`o}}, {Treu},
  {Nanayakkara}, {Jones}, {Merlin}, {Abraham}, {Stevens}, {Vulcani}, {Yang},
  {Bonchi}, {Bradac}, {Castellano}, {Fontana}, {Mason}, {Morishita}, {Paris},
  {Trenti}, {Marchesini}, {Wang}, \& {Santini}}]{Jacobs2022}
{Jacobs}, C., {Glazebrook}, K., {Calabr{\`o}}, A., {et~al.} 2022, arXiv
  e-prints, arXiv:2208.06516.
\newblock \doarXiv{2208.06516}

\bibitem[{{Kuchinski} {et~al.}(2001){Kuchinski}, {Madore}, {Freedman}, \&
  {Trewhella}}]{kuchinski01}
{Kuchinski}, L.~E., {Madore}, B.~F., {Freedman}, W.~L., \& {Trewhella}, M.
  2001, \aj, 122, 729, \dodoi{10.1086/321181}

\bibitem[{{Law} {et~al.}(2012){Law}, {Steidel}, {Shapley}, {Nagy}, {Reddy}, \&
  {Erb}}]{law12}
{Law}, D.~R., {Steidel}, C.~C., {Shapley}, A.~E., {et~al.} 2012, \apj, 745, 85,
  \dodoi{10.1088/0004-637X/745/1/85}

\bibitem[{{Lee} {et~al.}(2013){Lee}, {Giavalisco}, {Williams}, {Guo}, {Lotz},
  {Van der Wel}, {Ferguson}, {Faber}, {Koekemoer}, {Grogin}, {Kocevski},
  {Conselice}, {Wuyts}, {Dekel}, {Kartaltepe}, \& {Bell}}]{Lee13}
{Lee}, B., {Giavalisco}, M., {Williams}, C.~C., {et~al.} 2013, \apj, 774, 47,
  \dodoi{10.1088/0004-637X/774/1/47}

\bibitem[{{Leethochawalit} {et~al.}(2022){Leethochawalit}, {Trenti}, {Santini},
  {Yang}, {Merlin}, {Castellano}, {Fontana}, {Treu}, {Mason}, {Glazebrook},
  {Jones}, {Vulcani}, {Nanayakkara}, {Marchesini}, {Mascia}, {Morishita},
  {Roberts-Borsani}, {Bonchi}, {Paris}, {Boyett}, {Strait}, {Calabro`},
  {Pentericci}, {Bradac}, {Wang}, \& {Scarlata}}]{Leethochawalit2022}
{Leethochawalit}, N., {Trenti}, M., {Santini}, P., {et~al.} 2022, ApJ,
  submitted, arXiv:2207.11135.
\newblock \doarXiv{2207.11135}

\bibitem[{{Lotz} {et~al.}(2004){Lotz}, {Primack}, \& {Madau}}]{lotz04}
{Lotz}, J.~M., {Primack}, J., \& {Madau}, P. 2004, \aj, 128, 163,
  \dodoi{10.1086/421849}

\bibitem[{{Merlin} {et~al.}(2022){Merlin}, {Bonchi}, {Paris}, {Belfiori},
  {Fontana}, {Castellano}, {Nonino}, {Polenta}, {Santini}, {Yang},
  {Glazebrook}, {Treu}, {Roberts-Borsani}, {Trenti}, {Birrer}, {Brammer},
  {Grillo}, {Calabr{\`o}}, {Marchesini}, {Mason}, {Mercurio}, {Morishita},
  {Strait}, {Boyett}, {Leethochawalit}, {Nanayakkara}, {Vulcani}, {Bradac}, \&
  {Wang}}]{Merlin2022}
{Merlin}, E., {Bonchi}, A., {Paris}, D., {et~al.} 2022, arXiv e-prints,
  arXiv:2207.11701.
\newblock \doarXiv{2207.11701}

\bibitem[{{Morishita} {et~al.}(2014){Morishita}, {Ichikawa}, \&
  {Kajisawa}}]{morishita14}
{Morishita}, T., {Ichikawa}, T., \& {Kajisawa}, M. 2014, \apj, 785, 18,
  \dodoi{10.1088/0004-637X/785/1/18}

\bibitem[{{Oesch} {et~al.}(2010){Oesch}, {Bouwens}, {Carollo}, {Illingworth},
  {Trenti}, {Stiavelli}, {Magee}, {Labb{\'e}}, \& {Franx}}]{oesch10}
{Oesch}, P.~A., {Bouwens}, R.~J., {Carollo}, C.~M., {et~al.} 2010, \apjl, 709,
  L21, \dodoi{10.1088/2041-8205/709/1/L21}

\bibitem[{{Papovich} {et~al.}(2005){Papovich}, {Dickinson}, {Giavalisco},
  {Conselice}, \& {Ferguson}}]{papovich05}
{Papovich}, C., {Dickinson}, M., {Giavalisco}, M., {Conselice}, C.~J., \&
  {Ferguson}, H.~C. 2005, \apj, 631, 101, \dodoi{10.1086/429120}

\bibitem[{{Pawlik} {et~al.}(2016){Pawlik}, {Wild}, {Walcher}, {Johansson},
  {Villforth}, {Rowlands}, {Mendez-Abreu}, \& {Hewlett}}]{pawlik16}
{Pawlik}, M.~M., {Wild}, V., {Walcher}, C.~J., {et~al.} 2016, \mnras, 456,
  3032, \dodoi{10.1093/mnras/stv2878}

\bibitem[{{Rawat} {et~al.}(2009){Rawat}, {Wadadekar}, \& {De Mello}}]{rawat09}
{Rawat}, A., {Wadadekar}, Y., \& {De Mello}, D. 2009, \apj, 695, 1315,
  \dodoi{10.1088/0004-637X/695/2/1315}

\bibitem[{{Ribeiro} {et~al.}(2016){Ribeiro}, {Le F{\`e}vre}, {Tasca}, {Lemaux},
  {Cassata}, {Garilli}, {Maccagni}, {Zamorani}, {Zucca}, {Amor{\'\i}n},
  {Bardelli}, {Fontana}, {Giavalisco}, {Hathi}, {Koekemoer}, {Pforr}, {Tresse},
  \& {Dunlop}}]{ribeiro16}
{Ribeiro}, B., {Le F{\`e}vre}, O., {Tasca}, L.~A.~M., {et~al.} 2016, \aap, 593,
  A22, \dodoi{10.1051/0004-6361/201628249}

\bibitem[{{Rieke} {et~al.}(2005){Rieke}, {Kelly}, \& {Horner}}]{NIRCAM}
{Rieke}, M.~J., {Kelly}, D., \& {Horner}, S. 2005, in Society of Photo-Optical
  Instrumentation Engineers (SPIE) Conference Series, Vol. 5904, Cryogenic
  Optical Systems and Instruments XI, ed. J.~B. {Heaney} \& L.~G. {Burriesci},
  1--8, \dodoi{10.1117/12.615554}

\bibitem[{{Roberts-Borsani} {et~al.}(2022){Roberts-Borsani}, {Morishita},
  {Treu}, {Brammer}, {Strait}, {Wang}, {Bradac}, {Acebron}, {Bergamini},
  {Boyett}, {Calabr{\'o}}, {Castellano}, {Fontana}, {Glazebrook}, {Grillo},
  {Henry}, {Jones}, {Malkan}, {Marchesini}, {Mascia}, {Mason}, {Mercurio},
  {Merlin}, {Nanayakkara}, {Pentericci}, {Rosati}, {Santini}, {Scarlata},
  {Trenti}, {Vanzella}, {Vulcani}, \& {Willott}}]{RobertsBorsani2022}
{Roberts-Borsani}, G., {Morishita}, T., {Treu}, T., {et~al.} 2022, arXiv
  e-prints, arXiv:2207.11387.
\newblock \doarXiv{2207.11387}

\bibitem[{{Shibuya} {et~al.}(2015){Shibuya}, {Ouchi}, \&
  {Harikane}}]{shibuya15}
{Shibuya}, T., {Ouchi}, M., \& {Harikane}, Y. 2015, \apjs, 219, 15,
  \dodoi{10.1088/0067-0049/219/2/15}

\bibitem[{{Treu} {et~al.}(2022){Treu}, {Roberts-Borsani}, {Bradac}, {Brammer},
  {Fontana}, {Henry}, {Mason}, {Morishita}, {Pentericci}, {Wang}, {Acebron},
  {Bagley}, {Bergamini}, {Belfiori}, {Bonchi}, {Boyett}, {Boutsia},
  {Calabr{\'o}}, {Caminha}, {Castellano}, {Dressler}, {Glazebrook}, {Grillo},
  {Jacobs}, {Jones}, {Kelly}, {Leethochawalit}, {Malkan}, {Marchesini},
  {Mascia}, {Mercurio}, {Merlin}, {Nanayakkara}, {Nonino}, {Paris},
  {Poggianti}, {Rosati}, {Santini}, {Scarlata}, {Shipley}, {Strait}, {Trenti},
  {Tubthong}, {Vanzella}, {Vulcani}, \& {Yang}}]{TreuGlass22}
{Treu}, T., {Roberts-Borsani}, G., {Bradac}, M., {et~al.} 2022, \apj, 935, 110,
  \dodoi{10.3847/1538-4357/ac8158}

\bibitem[{{van der Wel} {et~al.}(2014){van der Wel}, {Franx}, {van Dokkum},
  {Skelton}, {Momcheva}, {Whitaker}, {Brammer}, {Bell}, {Rix}, {Wuyts},
  {Ferguson}, {Holden}, {Barro}, {Koekemoer}, {Chang}, {McGrath},
  {H{\"a}ussler}, {Dekel}, {Behroozi}, {Fumagalli}, {Leja}, {Lundgren},
  {Maseda}, {Nelson}, {Wake}, {Patel}, {Labb{\'e}}, {Faber}, {Grogin}, \&
  {Kocevski}}]{vanderwel14}
{van der Wel}, A., {Franx}, M., {van Dokkum}, P.~G., {et~al.} 2014, \apj, 788,
  28, \dodoi{10.1088/0004-637X/788/1/28}

\bibitem[{{Whitney} {et~al.}(2021){Whitney}, {Ferreira}, {Conselice}, \&
  {Duncan}}]{whitney21}
{Whitney}, A., {Ferreira}, L., {Conselice}, C.~J., \& {Duncan}, K. 2021, \apj,
  919, 139, \dodoi{10.3847/1538-4357/ac1422}

\bibitem[{{Wuyts} {et~al.}(2012){Wuyts}, {F{\"o}rster Schreiber}, {Genzel},
  {Guo}, {Barro}, {Bell}, {Dekel}, {Faber}, {Ferguson}, {Giavalisco}, {Grogin},
  {Hathi}, {Huang}, {Kocevski}, {Koekemoer}, {Koo}, {Lotz}, {Lutz}, {McGrath},
  {Newman}, {Rosario}, {Saintonge}, {Tacconi}, {Weiner}, \& {van der
  Wel}}]{wuyts12}
{Wuyts}, S., {F{\"o}rster Schreiber}, N.~M., {Genzel}, R., {et~al.} 2012, \apj,
  753, 114, \dodoi{10.1088/0004-637X/753/2/114}

\bibitem[{{Yang} {et~al.}(2022{\natexlab{a}}){Yang}, {Leethochawalit}, {Treu},
  {Roberts-Borsani}, {Brada{\v{c}}}, {Birrer}, {Castellano}, {Merlin},
  {Fontana}, {Amorin}, \& {Trenti}}]{Yang22}
{Yang}, L., {Leethochawalit}, N., {Treu}, T., {et~al.} 2022{\natexlab{a}},
  arXiv e-prints, arXiv:2201.08858.
\newblock \doarXiv{2201.08858}

\bibitem[{{Yang} {et~al.}(2022{\natexlab{b}}){Yang}, {Morishita},
  {Leethochawalit}, {Castellano}, {Calabro}, {Treu}, {Bonchi}, {Fontana},
  {Mason}, {Merlin}, {Paris}, {Trenti}, {Roberts-Borsani}, {Bradac},
  {Vanzella}, {Vulcani}, {Marchesini}, {Ding}, {Nanayakkara}, {Birrer},
  {Glazebrook}, {Jones}, {Boyett}, {Santini}, {Strait}, \& {Wang}}]{Yang22b}
{Yang}, L., {Morishita}, T., {Leethochawalit}, N., {et~al.} 2022{\natexlab{b}},
  arXiv e-prints, arXiv:2207.13101.
\newblock \doarXiv{2207.13101}

\end{thebibliography}

\clearpage

\begin{sidewaystable*}
\rule{-3.5cm}{+1.501cm} 
\resizebox{1\textwidth}{!}{%
\centering
\renewcommand{\arraystretch}{1}
\begin{tabular}{|l|l|l|l|l|l|l|l|l|l|l|l|l|l|l|l|l|l|l|l|} 
\hline

\multicolumn{2}{l}{\textbf{\large{ }}} & \multicolumn{6}{l}{\textbf{\large{ F115W}}} & \multicolumn{6}{l}{\textbf{\large{F150W}}} & \multicolumn{6}{l}{\textbf{\large{F200W}}} \\ 

\hline
ID & z$_{phot}$ & SNpix & gini & M$_{20}$ & C & A$_S$ & S & SNpix & gini & M$_{20}$ & C & A$_S$ & S & SNpix & gini & M$_{20}$ & C & A$_S$ & S \\ 
\hline
1 & 10.74 & -9.0 & -9.0 & -9.0 & -9.0 & -9.0 & -9.0 & 2.8 & 0.583 & -1.582 & 2.378 & 0.003 & 0.107 & 3.6 & 0.543 & -1.743 & 1.923 & 0.047 & 0.109 \\
2 & 12.3 & -9.0 & -9.0 & -9.0 & -9.0 & -9.0 & -9.0 & -9.0 & -9.0 & -9.0 & -9.0 & -9.0 & -9.0 & 3.7 & 0.605 & -1.481 & 3.569 & 0.052 & 0.063 \\
3 & 11.1478 & -9.0 & -9.0 & -9.0 & -9.0 & -9.0 & -9.0 & 2.9 & 0.496 & -1.237 & 2.081 & 0.097 & 0.108 & 3.1 & 0.494 & -1.171 & 2.032 & 0.098 & 0.079 \\
4 & 10.0814 & -9.0 & -9.0 & -9.0 & -9.0 & -9.0 & -9.0 & 3.2 & 0.489 & -0.815 & 1.717 & 0.156 & 0.151 & 3.8 & 0.473 & -0.734 & 1.95 & 0.153 & 0.182 \\
5 & 9.4055 & -9.0 & -9.0 & -9.0 & -9.0 & -9.0 & -9.0 & -9.0 & -9.0 & -9.0 & -9.0 & -9.0 & -9.0 & 2.7 & 0.497 & -1.001 & 1.725 & -0.046 & 0.148 \\
6 & 9.9013 & -9.0 & -9.0 & -9.0 & -9.0 & -9.0 & -9.0 & 1.5 & -9.0 & -9.0 & -9.0 & -9.0 & -9.0 & 2.6 & 0.465 & -0.915 & 1.299 & 0.091 & 0.089 \\
1470 & 7.6 & 1.8 & -9.0 & -9.0 & -9.0 & -9.0 & -9.0 & -9.0 & -9.0 & -9.0 & -9.0 & -9.0 & -9.0 & 3.5 & 0.46 & -1.252 & 1.027 & 0.653 & 0.2 \\
2236 & 8.0 & 2.7 & 0.54 & -1.638 & 2.151 & 0.052 & 0.065 & 2.9 & 0.587 & -1.543 & 2.022 & -0.02 & 0.058 & 2.4 & 0.526 & -1.867 & 2.923 & 0.059 & 0.084 \\
2574 & 7.4 & 2.9 & -9.0 & -9.0 & 1.886 & -9.0 & -9.0 & 3.1 & 0.501 & -1.36 & 2.096 & 0.102 & 0.106 & 2.9 & 0.527 & -1.722 & 2.086 & 0.053 & 0.065 \\
2911$^\ast$ & 6.9 & 2.9 & 0.562 & -0.961 & 1.18 & 0.29 & 0.182 & 3.4 & 0.519 & -1.058 & 0.894 & 0.181 & 0.209 & 3.7 & 0.516 & -1.197 & 0.935 & 0.202 & 0.211 \\
3120 & 7.4 & 3.9 & 0.462 & -0.796 & 1.567 & 0.135 & -0.009 & 4.5 & 0.459 & -0.767 & 1.599 & 0.102 & 0.145 & 4.1 & 0.478 & -0.791 & 1.461 & 0.034 & 0.088 \\
4542 & 9.0 & 1.7 & -9.0 & -9.0 & -9.0 & -9.0 & -9.0 & 2.5 & 0.603 & -1.723 & 2.73 & 0.056 & 0.07 & 3.6 & 0.491 & -1.393 & 2.572 & 0.134 & 0.088 \\
4863 & 8.1 & 3.0 & 0.495 & -1.182 & 2.167 & 0.112 & 0.054 & 2.5 & 0.563 & -1.486 & 2.513 & 0.018 & 0.037 & 3.9 & 0.486 & -1.635 & 2.805 & 0.052 & 0.088 \\
5001 & 8.1 & -9.0 & -9.0 & -9.0 & -9.0 & -9.0 & -9.0 & -9.0 & -9.0 & -9.0 & -9.0 & -9.0 & -9.0 & -9.0 & -9.0 & -9.0 & -9.0 & -9.0 & -9.0 \\
1708 & 7.8 & 3.3 & 0.517 & -1.138 & 1.95 & -0.004 & 0.09 & -9.0 & -9.0 & -9.0 & -9.0 & -9.0 & -9.0 & 3.3 & 0.524 & -1.647 & 2.667 & 0.099 & 0.061 \\
4397 & 8.1 & 2.3 & 0.542 & -1.072 & 1.531 & 0.105 & 0.15 & -9.0 & -9.0 & -9.0 & -9.0 & -9.0 & -9.0 & 2.9 & 0.508 & -0.962 & 1.904 & 0.065 & 0.156 \\
6116 & 8.2 & -9.0 & -9.0 & -9.0 & -9.0 & -9.0 & -9.0 & -9.0 & -9.0 & -9.0 & -9.0 & -9.0 & -9.0 & -9.0 & -9.0 & -9.0 & -9.0 & -9.0 & -9.0 \\
6263 & 8.2 & -9.0 & -9.0 & -9.0 & -9.0 & -9.0 & -9.0 & -9.0 & -9.0 & -9.0 & -9.0 & -9.0 & -9.0 & 2.5 & 0.529 & -1.321 & 2.173 & 0.027 & 0.075 \\

\hline

\multicolumn{2}{l}{\textbf{\large{ }}} & \multicolumn{6}{l}{\textbf{\large{ F277W}}} & \multicolumn{6}{l}{\textbf{\large{F356W}}} & \multicolumn{6}{l}{\textbf{\large{F444W}}} \\ 

\hline
ID & z$_{phot}$ & SNpix & gini & M$_{20}$ & C & A$_S$ & S & SNpix & gini & M$_{20}$ & C & A$_S$ & S & SNpix & gini & M$_{20}$ & C & A$_S$ & S \\  
\hline
1 & 10.74 & 6.2 & 0.482 & -1.534 & 2.48 & 0.068 & 0.108 & 4.6 & 0.596 & -1.764 & 2.736 & 0.124 & 0.077 & 6.2 & 0.515 & -1.761 & 2.338 & 0.052 & 0.044 \\
2 & 12.3 & 6.0 & 0.525 & -1.019 & 2.094 & 0.07 & 0.068 & 5.6 & 0.491 & -1.237 & 2.339 & 0.07 & 0.032 & 6.0 & 0.499 & -1.416 & 2.051 & 0.029 & 0.095 \\
3 & 11.1478 & 5.6 & 0.499 & -1.387 & 2.067 & 0.006 & 0.064 & 4.6 & 0.529 & -1.652 & 2.112 & 0.035 & 0.038 & 4.8 & 0.519 & -1.434 & 2.814 & 0.089 & 0.056 \\
4 & 10.0814 & 4.4 & 0.464 & -0.933 & 2.726 & 0.041 & 0.053 & 3.2 & 0.473 & -0.951 & 1.88 & 0.07 & 0.115 & 3.9 & 0.441 & -0.883 & 2.142 & 0.114 & 0.085 \\
5 & 9.4055 & 6.1 & 0.435 & -0.844 & 1.818 & 0.358 & 0.186 & 5.9 & 0.422 & -0.861 & 1.481 & 0.073 & 0.152 & 6.0 & 0.429 & -0.868 & 2.39 & 0.063 & 0.166 \\
6 & 9.9013 & 4.1 & 0.466 & -0.976 & 0.986 & 0.174 & 0.115 & 4.3 & 0.475 & -1.004 & 1.167 & 0.187 & 0.123 & 5.0 & 0.493 & -1.015 & 1.49 & 0.166 & 0.14 \\
1470 & 7.6 & 4.2 & 0.507 & -1.165 & 1.06 & 0.477 & 0.222 & 4.9 & 0.482 & -0.954 & 1.276 & 0.349 & 0.181 & 5.2 & 0.48 & -0.827 & 1.29 & 0.342 & 0.195 \\
2236 & 8.0 & 4.6 & 0.416 & -1.091 & 2.252 & 0.038 & 0.121 & 5.2 & 0.385 & -1.094 & 2.144 & 0.164 & 0.047 & 4.7 & 0.498 & -1.568 & 2.025 & 0.081 & 0.004 \\
2574 & 7.4 & 4.4 & 0.458 & -1.292 & 1.318 & 0.046 & 0.04 & 4.5 & 0.491 & -1.383 & 2.938 & 0.198 & 0.102 & 4.9 & 0.523 & -1.679 & 2.356 & 0.089 & 0.07 \\
2911$^\ast$ & 6.9 & 5.5 & 0.549 & -0.943 & 1.136 & 0.209 & 0.262 & 7.3 & 0.547 & -1.074 & 1.136 & 0.16 & 0.316 & 6.7 & 0.547 & -1.097 & 1.154 & 0.085 & 0.31 \\
3120 & 7.4 & 7.1 & 0.449 & -1.593 & 1.85 & 0.09 & 0.034 & 8.0 & 0.459 & -1.384 & 1.587 & 0.113 & 0.058 & 7.9 & 0.429 & -1.091 & 1.663 & 0.118 & 0.037 \\
4542 & 9.0 & 4.7 & 0.466 & -1.456 & 2.291 & 0.058 & 0.071 & 4.3 & 0.423 & -1.693 & 1.806 & 0.148 & 0.061 & 4.1 & 0.529 & -1.712 & 1.897 & 0.059 & 0.108 \\
4863 & 8.1 & 4.4 & 0.48 & -0.808 & 2.009 & 0.066 & 0.058 & 4.6 & 0.501 & -1.297 & 1.594 & 0.061 & 0.031 & 4.9 & 0.531 & -1.648 & 2.41 & 0.175 & 0.103 \\
5001 & 8.1 & 4.2 & 0.405 & -1.17 & 1.384 & 0.082 & 0.08 & 4.0 & 0.456 & -0.799 & 2.112 & 0.137 & 0.155 & 4.1 & 0.463 & -1.198 & 1.743 & 0.081 & 0.099 \\
1708 & 7.8 & 4.6 & 0.491 & -1.368 & 3.008 & 0.208 & 0.138 & -9.0 & -9.0 & -9.0 & -9.0 & -9.0 & -9.0 & 7.3 & 0.515 & -1.468 & 2.74 & 0.052 & 0.133 \\
4397 & 8.1 & 4.8 & 0.464 & -1.927 & 2.217 & 0.049 & 0.128 & -9.0 & -9.0 & -9.0 & -9.0 & -9.0 & -9.0 & 5.1 & 0.531 & -1.667 & 2.394 & 0.026 & 0.096 \\
6116 & 8.2 & 4.2 & 0.478 & -0.716 & 1.454 & 0.073 & 0.125 & 5.3 & 0.442 & -0.997 & 1.86 & 0.189 & 0.033 & 6.7 & 0.471 & -1.421 & 2.035 & 0.059 & 0.03 \\
6263 & 8.2 & 4.5 & 0.486 & -1.168 & 1.589 & 0.082 & 0.095 & -9.0 & -9.0 & -9.0 & -9.0 & -9.0 & -9.0 & 5.3 & 0.523 & -1.286 & 2.432 & 0.135 & 0.079 \\

\hline
\end{tabular} 
}

\hfill
\caption{\large Table with the morphological parameters estimated for our sample. $^\ast$ = This measurement actually refers to the two interacting galaxies ID $2911$ and $2936$ in the catalog of \citet{Leethochawalit2022}.  }\label{tab:parameters}
\end{sidewaystable*}



\end{document}